\documentclass[12pt]{article}

\usepackage{graphicx}
\usepackage{amsmath}
\usepackage{amssymb}
\usepackage{hyperref}

\begin{document}

\title{Investigating the Ground Energy Distribution of Particles Produced in Extensive Air Showers}
\author{ITAB F. HUSSEIN and Al-RUBAIEE A. A.\\
Department of Physics, College of Science, Mustansiriyah University, Baghdad, Iraq\\
\texttt{itabfadhil@uomustansiriyah.edu.iq, dr.rubaiee@uomustansiriyah.edu.iq}}
\date{}

\maketitle

\begin{abstract}
The energy spectra of particles arriving at the ground is a significant observable in the analysis of extensive air showers (EAS). Energy distributions at ground were studied for primary particles ($^{12}$C, $^{56}$Fe, p, and $^{28}$Si) with high primary energies ($10^{17}$, $10^{18}$, $10^{19}$, and $10^{20}$ eV) from two zenith angles ($0^{\circ}$ and $30^{\circ}$). 960 EAS were simulated using the Monte-Carlo program Aires (version 19.04.00) with three models of hadronic interaction (EPOS-LHC, QGSJET-II-04, and Sibyll2.3c). Good agreement was obtained by comparing the present results with results simulated using CORSIKA for primary iron at an energy of $10^{20}$ eV. In this study we investigated various secondary particles that arrive at the ground and deposit a portion of their energy on ground detectors. These results show that the distinction in energy distribution at ground is greater for primary protons than carbon, iron, or silicon nuclei at higher energies and steeper zenith angles.
\end{abstract}

\section{Introduction}
When ultra-high energy cosmic rays (UHECRs) enter the Earth's atmosphere, they initially interact with oxygen or nitrogen molecules in the air, resulting in complex interactions and cascades that produce extensive air showers (EAS) containing hundreds of trillions of particles \cite{Blumer2009, Knapp2003, Dongsu2011}. Original particle characteristics, such as energy, direction of arrival, and element are derived by detection of secondary particles that reach the ground. Primary particles, which collide with the ground at high energies, are not directly detectable. A cascade of particles is released when they collide with the atmosphere, which is detected by telescopes and ground equipment. From the detected shower indicator, the characteristics of this primary particle can be recreated \cite{Aab2014, Bellido2017, ALRubaiee2021}.

In EAS, only a small percentage of secondary particles make it to the ground. Ground detectors, such as water Cherenkov tanks or scintillation detectors, collect a portion of the energy emitted by these particles \cite{Hillas1971}. Ionization and bremsstrahlung are two processes that cause electrons and muons to lose energy. Ionization is the primary source of energy loss for muons. Bremsstrahlung does not cause significant energy losses until muon energies in the thousands of GeV are reached. For electrons, on the other hand, bremsstrahlung does result in energy loss for particles with moderate starting energies \cite{Sciuttu2002}.

Models of hadronic interaction play a significant role in the estimation of EAS features. Alternatively, different models used in the AIRES simulation code have introduced phenomenological methods. In this work, the energy distribution at ground ($E_{\text{ground}}$) (i.e., the total energy deducted from the rest-mass energy) \cite{Drescher2003} for electrons, muons and pions were studied, by several models of hadronic interaction usually applied to air shower simulation, EPOS-LHC \cite{Pierog2015}, QGSJetII.04 \cite{Ostapchenko2011} and Sibyll2.3c \cite{Ahn2009}. These models have the best representation of high-energy hadronic interactions \cite{Klages1997}.

\section{Energy Estimation via Heitler and Matthews Models}
On a microscopic level, simple cascade models are offering some insight into the relationship between air shower observables and interaction physics \cite{Matthews2001, Pierog2006, Heitler1954}. Heitler's model of particle cascades can be used to define the major aspects of electromagnetic shower profiles \cite{Matthews2005, AlvarezMuniz2002}. Assume that a particle (electron, positron, or photon) divides its energy ($E_0$) evenly into two separate particles, after traveling $X_0$ radiation length in the air and allows secondary particles to frequent this process as shown in the Figure \ref{fig:heitler_models}:

\begin{figure}[htbp]
\centering
\includegraphics[width=0.8\textwidth]{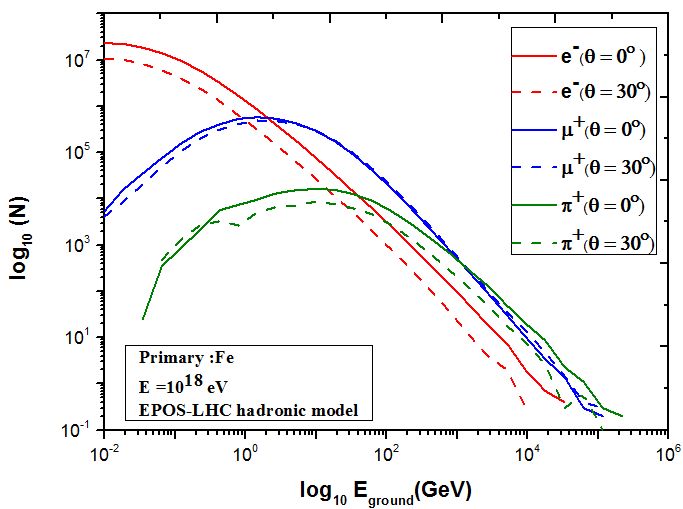}
\caption{Diagrammatic views of (a) an electromagnetic Heitler model \& (b) Hadronic Heitler-Matthews model \cite{Matthews2001, Aartsen2013}.}
\label{fig:heitler_models}
\end{figure}

We get a particle cascade after $n$ radiation durations that has included into ($N = 2^n$) and energy equal ($E = E_0/N$). Multiplication stops, when the particle energies are too low for pair production or bremsstrahlung. This energy is referred to the critical energy ($\varepsilon_c$). At this point, maximum particle number is attained, known as, ($N_{\text{max}}$) when the energy of all particles is the same, then:

\begin{equation}
N_{\text{max}} = \frac{E_0}{\varepsilon_c}
\end{equation}

Since
\begin{equation}
N_{\text{max}} = 2^{n_{\text{max}}}
\end{equation}

Where
\begin{equation}
n_{\text{max}} = \frac{\ln(E_0/\varepsilon_c)}{\ln 2}
\end{equation}

From Eq. (3) $N_{\text{max}}$ is directly proportional to the primary energy $E_0$ \cite{Linsley1977}. EAS was first modeled on protons by Matthews following a method similar to Heitler's, charged pions $\pi^{\pm}$ and neutral pions $\pi^0$ are produced when protons traverse one interaction length and interact, which decays into photons, immediately begin an electromagnetic shower. As for the electromagnetic cascade, during the particle production, we assume the same energy split. Following $n$ interactions:

\begin{equation}
N_{\pi} = (N_{\text{mult}})^n
\end{equation}

The total energy of the charged pions produced is $E_{\pi^{\pm}} = \frac{2}{3}E_0$. After $n$ interactions, the energy per charged pion is:

\begin{equation}
E_{\pi^{\pm}} = \frac{2E_0}{3(N_{\text{mult}})^n}
\end{equation}

The process end When the energy of pions falls under the critical energy $\varepsilon_c^{\pi}$, they decay into muons. The muons number is $N_{\mu} = N_{\pi}$, where $n_c$ is the number of length interaction needed to exceed the interaction length of the charged pion:

\begin{equation}
n_c = \frac{\ln(E_0/\varepsilon_c^{\pi})}{\ln(N_{\text{mult}})}
\end{equation}

Therefore, the entire energy is split into two electromagnetic and hadronic channels
\begin{equation}
E_0 = E_{\text{em}} + E_{\text{had}}
\end{equation}

The muon number is thus reliant on the secondary hadronic abundance and pion charge ratio. According to Matthews model, energy is provided by a linear combination of the electron and muon sizes. This finding is unaffected by transitions in energy separated between the electromagnetic and hadronic channels, and it is unaffected by the parent particle's mass \cite{Bergmann2007}.

\section{The Simulation of EAS using AIRES System}
Extensive shower simulations using the program AIRES "AIR-shower Extended Simulations" version (19.04.00) is a Monte-Carlo simulation program. There were four atomic nuclei to consider: carbon, iron, proton and silicon with energies ($10^{17}$, $10^{18}$, $10^{19}$ and $10^{20}$) eV with zenith angles ($0^{\circ}$ and $30^{\circ}$). At the level of the ground 1400 m above the equivalent sea level to slant depth 1000 g/cm$^2$. Cut energies for gamma, electrons, muons and meson are 80 KeV, 80 KeV, 10 MeV and 60 MeV, respectively. Also the energy of thinning algorithm was set to ($\varepsilon_{\text{th}}=10^{-6}$), in addition the effects of three models of hadronic interaction were used: QGSJetII.04, EPOS-LHC and Sibyll2.3c on the energy distribution at ground of secondary charged particles produced in the EAS is taken into consideration.

\section{Results and Discussion}
\begin{figure}[htbp]
\centering
\includegraphics[width=0.8\textwidth]{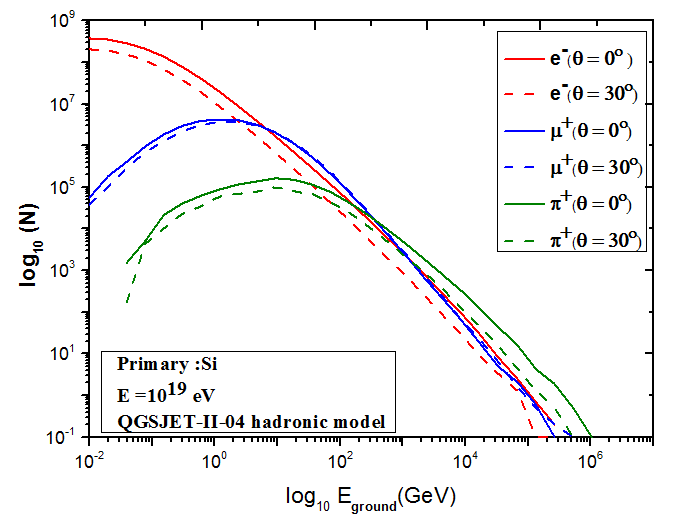}
\caption{The number of secondary particles as a function of the energy distribution at ground for various primary particles and various energies for: vertical showers (solid lines) and inclined showers (dashed lines) using the EPOS-LHC hadronic model.}
\label{fig:epos_model}
\end{figure}

\begin{figure}[htbp]
\centering
\includegraphics[width=0.8\textwidth]{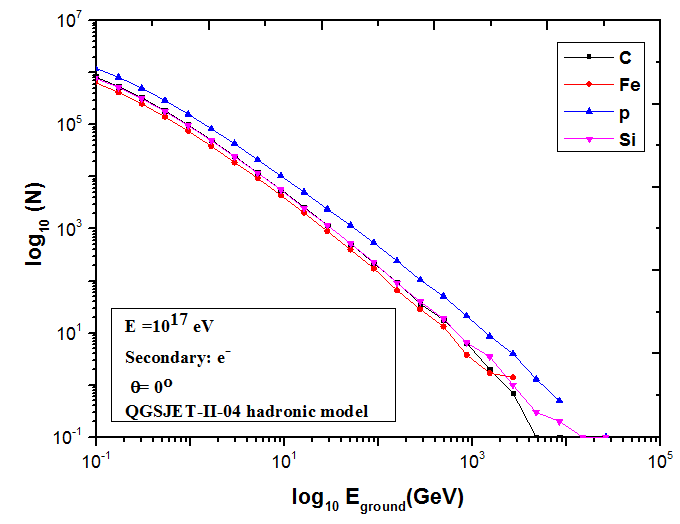}
\caption{The number of secondary particles as a function of the energy distribution at ground for various primary particles and various energies for: vertical showers (solid lines) and inclined showers (dashed lines) using the QGSJET-II-04 hadronic model.}
\label{fig:qgsjet_model}
\end{figure}

\begin{figure}[htbp]
\centering
\includegraphics[width=0.8\textwidth]{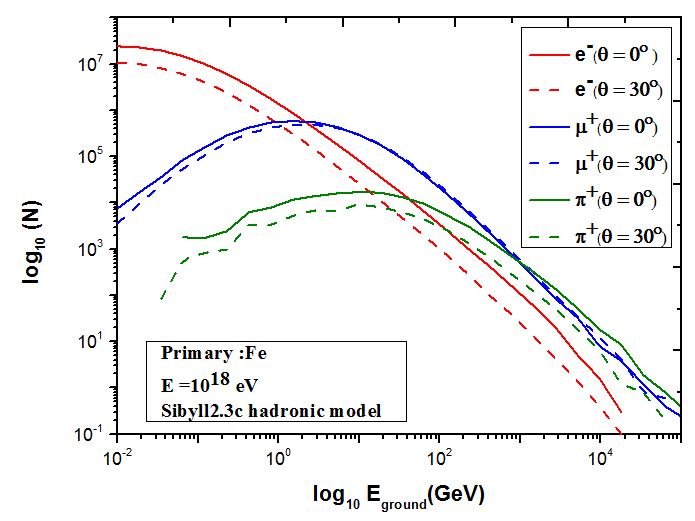}
\caption{The number of secondary particles as a function of the energy distribution at ground for various primary particles and various energies for: vertical showers (solid lines) and inclined showers (dashed lines) using the Sibyll2.3c hadronic model.}
\label{fig:sibyll_model}
\end{figure}

Secondary particles, like electrons, muons, and pions, carry the vast majority of the energy in EAS to the ground. Figures \ref{fig:epos_model}, \ref{fig:qgsjet_model}, and \ref{fig:sibyll_model} show particles number as a function of energy distribution of secondary particles at the ground in EAS of (C, Fe, p, and Si) primaries with energies ($10^{17}$, $10^{18}$, $10^{19}$, and $10^{20}$) eV, and zenith angles of $0^{\circ}$ and $30^{\circ}$ simulated using EPOS-LHC, QGSJET-II-04 and Sibyll2.3c hadronic models respectively.

\begin{figure}[htbp]
\centering
\includegraphics[width=0.8\textwidth]{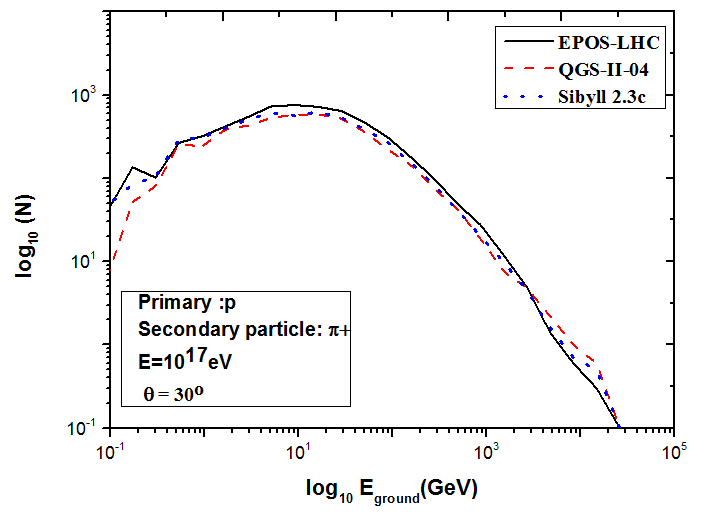}
\caption{The energy distribution at ground for secondary particles produced in EAS initiated by carbon primaries with different energies and zenith angles using the EPOS-LHC hadronic model.}
\label{fig:carbon_model}
\end{figure}

\begin{figure}[htbp]
\centering
\includegraphics[width=0.8\textwidth]{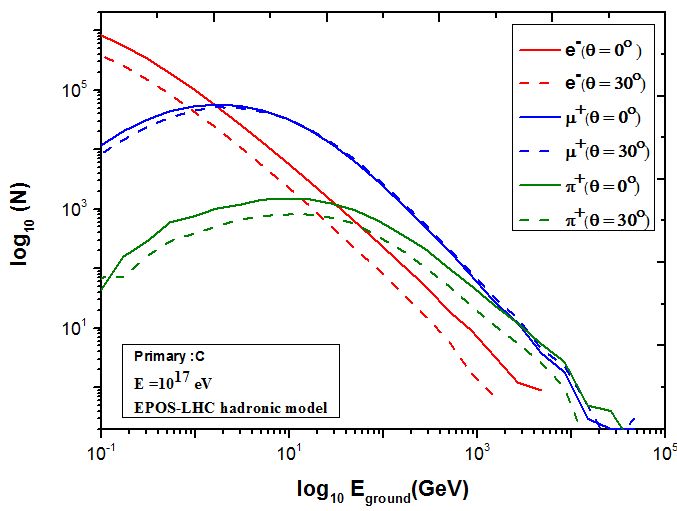}
\caption{The energy distribution at ground for secondary particles produced in EAS initiated by iron primaries with different energies and zenith angles using the EPOS-LHC hadronic model.}
\label{fig:iron_model}
\end{figure}

\begin{figure}[htbp]
\centering
\includegraphics[width=0.8\textwidth]{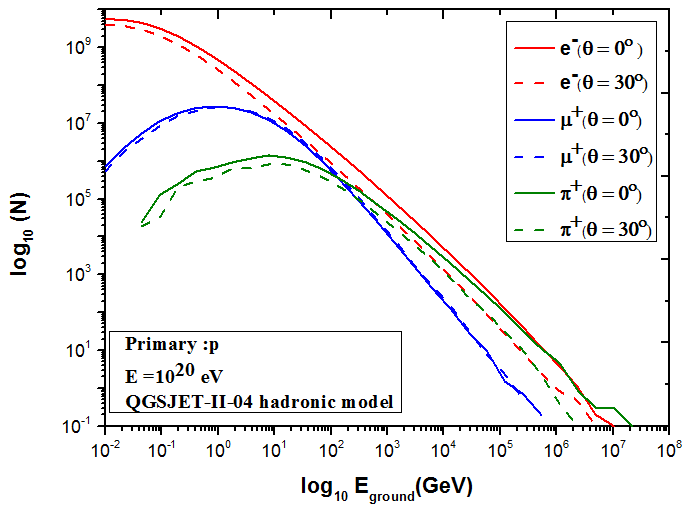}
\caption{The energy distribution at ground for secondary particles produced in EAS initiated by proton primaries with different energies and zenith angles using the EPOS-LHC hadronic model.}
\label{fig:proton_model}
\end{figure}

\begin{figure}[htbp]
\centering
\includegraphics[width=0.8\textwidth]{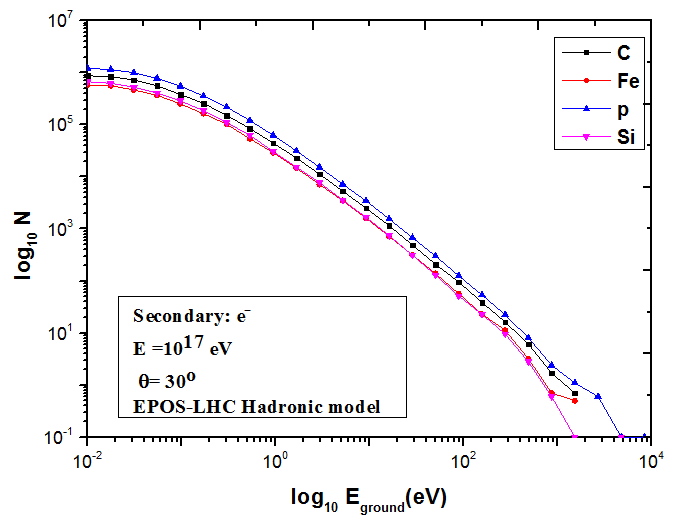}
\caption{The energy distribution at ground for secondary particles produced in EAS initiated by silicon primaries with different energies and zenith angles using the EPOS-LHC hadronic model.}
\label{fig:silicon_model}
\end{figure}

The energy distribution at ground for secondary particles produced in EAS was studied for different primary particles, energies, and zenith angles. Figures \ref{fig:carbon_model}, \ref{fig:iron_model}, \ref{fig:proton_model}, and \ref{fig:silicon_model} show the energy distribution at ground for secondary particles produced in EAS initiated by carbon, iron, proton, and silicon primaries, respectively, with energies ($10^{17}$, $10^{18}$, $10^{19}$, and $10^{20}$) eV and zenith angles of $0^{\circ}$ and $30^{\circ}$ using the EPOS-LHC hadronic model.

\section{Conclusions}
In this study, we investigated the energy distribution at ground for secondary particles produced in extensive air showers initiated by different primary particles (carbon, iron, proton, and silicon) with various energies ($10^{17}$, $10^{18}$, $10^{19}$, and $10^{20}$ eV) and zenith angles ($0^{\circ}$ and $30^{\circ}$) using different hadronic interaction models (EPOS-LHC, QGSJET-II-04, and Sibyll2.3c).

The results show that the distinction in energy distribution at ground is greater for primary protons than carbon, iron, or silicon nuclei at higher energies and steeper zenith angles. This is due to the fact that proton-initiated showers develop deeper in the atmosphere compared to heavier nuclei, resulting in more energetic particles reaching the ground.

The effect of hadronic interaction models on the energy distribution at ground was also studied, showing that the EPOS-LHC model predicts a higher number of particles with higher energies compared to the QGSJET-II-04 and Sibyll2.3c models. This is attributed to the different treatment of hadronic interactions in these models.

The comparison with CORSIKA simulations for iron primaries with energy $10^{20}$ eV shows good agreement, validating the results obtained using the AIRES simulation program.

These findings have important implications for the interpretation of experimental data from ground-based cosmic ray observatories, as they provide insights into the relationship between the energy distribution of particles at ground and the properties of the primary cosmic rays.


\end{document}